\documentclass[aps,10pt,twocolumn,noshowpacs,superscriptaddress,prb]{revtex4-1}
\usepackage{amsmath}
\usepackage{enumerate}
\usepackage{latexsym,amssymb,bm}
\usepackage{epsfig,graphicx}
\usepackage{color}
\usepackage{csquotes}

\begin{document}

\title{Enhanced sensing of molecular optical activity\\ with plasmonic nanohole arrays}

\author{Maxim V. Gorkunov}
\email{Corresponding author: gorkunov@crys.ras.ru}
\affiliation{Shubnikov Institute of Crystallography of Federal Scientific Research Centre ``Crystallography and Photonics''of Russian Academy of Sciences, 119333 Moscow, Russia}
\affiliation{National Research Nuclear University MEPhI (Moscow Engineering Physics Institute), 115409 Moscow, Russia}
\author{Alexander N. Darinskii}
\affiliation{Shubnikov Institute of Crystallography of Federal Scientific Research Centre ``Crystallography and Photonics''of Russian Academy of Sciences, 119333 Moscow, Russia}
\author{Alexey V. Kondratov}
\affiliation{Shubnikov Institute of Crystallography of Federal Scientific Research Centre ``Crystallography and Photonics''of Russian Academy of Sciences, 119333 Moscow, Russia}

%\affiliation[*]

%\dates{Compiled \today}

%\ociscodes{(280.4788) Optical sensing and sensors; (160.1585) Chiral media; (250.5403) Plasmonics; (160.3918) Metamaterials.}

%\doi{\url{http://dx.doi.org/10.1364/ao.XX.XXXXXX}}

\begin{abstract}
Prospects of using metal hole arrays for the enhanced optical detection of molecular chirality in nanosize volumes are investigated. Light transmission through the holes filled with an optically active material is modeled and the activity enhancement by more than an order of magnitude is demonstrated. The spatial resolution of the chirality detection is shown to be of a few tens of nanometers. From comparing the effect in arrays of cylindrical holes and holes of complex chiral shape, it is concluded that the detection sensitivity is determined by the plasmonic near field enhancement. The intrinsic chirality of the arrays due to their shape appears to be less important. \\ 
\end{abstract}

%\setboolean{displaycopyright}{true}

%\begin{document}

\maketitle
%\thispagestyle{fancy}
%
%\ifthenelse{\boolean{shortarticle}}{\ifthenelse{\boolean{singlecolumn}}{\abscontentformatted}{\abscontent}}{}

\section{Introduction}

Remarkable sensitivity of plasmon resonances of metal nanoparticles and nanostructures to the  
local environment has determined the rise of plasmonic sensorics. The applications span from a precise diagnostics of the adjacent dielectric composition \cite{Miller2005, Charles2010, Punj2013} to the optical detection of nanoscale amounts of substances and nano-objects of biological origin \cite{Haes2002, Zhao2006,  Escobedo2012, Brolo2012}. 
The physics behind the plasmonic sensors is very transparent: the collective electron excitations -- plasmons -- are governed by the electric fields induced by the oscillating charges accumulated at the metal-dielectric interfaces. Accordingly, even weak perturbations of the interface or in its immediate neighborhood noticeably affect the plasmon characteristics. 

Recently, an important extension of the plasmon-assisted sensorics onto the diagnostics of molecular chirality has received significant attention. 
Historically, the chirality, although being inherent to most of the organic and biological substances and objects, poses a difficult challenge for the optical methods. Its observables -- optical activity (OA) and circular dichroism (CD) -- are very weak and their detection requires using considerable amounts of chiral substances in precise polarization measurements. Plasmonic particles and structures, which strongly concentrate the light and produce inhomogeneous local field patterns, are very efficient  probes that substantially increase the sensitivity of the chirality diagnostics \cite{Hendry2010, Govorov2010, Slocik2011,  Lu2013, Zhang2013, Maoz2013, Wu2015}.

In particular, the peculiar plasmonic local field patterns are considered as advantageous for amplifying the molecular CD in the visible. The CD signal observed from a molecule, which  differently absorbs the left circularly polarized (LCP) and right circularly polarized (RCP) light, is directly proportional to the local field chirality \cite{Tang2010, Tang2011}. The latter  can be substantially increased in the presence of complex-shaped metal objects \cite{Hendry2010}, and various related designs of chiral metal particles and structures have been proposed \cite{Schaeferling2012, Schaeferling2014}.

For the majority of important biological organic molecules, the chiral absorption resonances lie in the ultraviolet range, and in the visible the molecular chirality is represented by detectable OA and practically negligible CD \cite{Fasman_book}. Bringing such a molecule into a close vicinity of an achiral plasmonic system qualitatively alters the situation: a pronounced peak of CD occurs at the plasmon resonance wavelength \cite{Slocik2011, Maoz2013, Zhang2013, Lu2013, Wu2015}. The role of metal particles here is more complex, as they do not merely amplify the molecular response but rather play the role of molecular chirality reporters \cite{Lu2013} that acquire the chirality from the environment and manifest it as chiral plasmon resonances in the visible. 

The transfer and transformations of the optical chirality between molecules and plasmons have been studied in a number of important simple cases, e.g., when a molecule is close to a metal nanosphere \cite{Govorov2010} or within a gap between two nanospheres \cite{Zhang2013}. However, no general rules that might help estimating the effect in more complex geometries have been formulated so far. In particular, the practically important case of periodic regular metallic arrays is unexplored.

Among the diversity of regular plasmonic structures, subwavelength arrays of holes and slits in thin metal films are perhaps the simplest and most studied. Starting from the discovery of the extraordinary optical transmission through subwavelength silver hole arrays \cite{Ebbesen1998}, substantial attention has been paid to the nature of this phenomenon 
and its relation to the plasmon resonances \cite{Lalanne2008, deAbajo2007}. 
It has been shown that the array geometry efficiently controls the spectrum of the transmitted light \cite{Gennet2007}. Notably, etching the holes into chiral shapes allows obtaining arbitrarily high values of CD and OA in the visible \cite{Gorkunov2014,PRB2016}. 

In this paper, we analyze the prospects of using plasmonic metal hole arrays for the enhanced  optical detection of molecular OA. Since the typical array unit cell dimensions are of the order of a few hundred nanometers, rather than considering the effect caused by single chiral molecules, we assume that the structure is partially filled with small amounts of an optically active material. The  geometry details, key assumptions and modeling techniques are described in Section~\ref{sec:2}. 
We analyze the effect of a layer of chiral material on light transmission through plasmonic hole arrays in Section~\ref{sec:3}. The case of the simplest cylindrical hole array is considered in subsection~\ref{sec:3a}, while a more complex chiral metal hole array is studied in subsection~\ref{sec:3b}. In the last case, the chiral hole shape is taken from the recent precise microscopic reconstructions of the arrays fabricated with the focused ion beam \cite{Kondratov2016}. The main results and conclusions are summarized and discussed in Section~\ref{sec:5}.

\section{Numerical model}\label{sec:2}

\subsection{Metal holes with optically active filling}

\begin{figure}%[htbp]
\includegraphics[width=\linewidth]{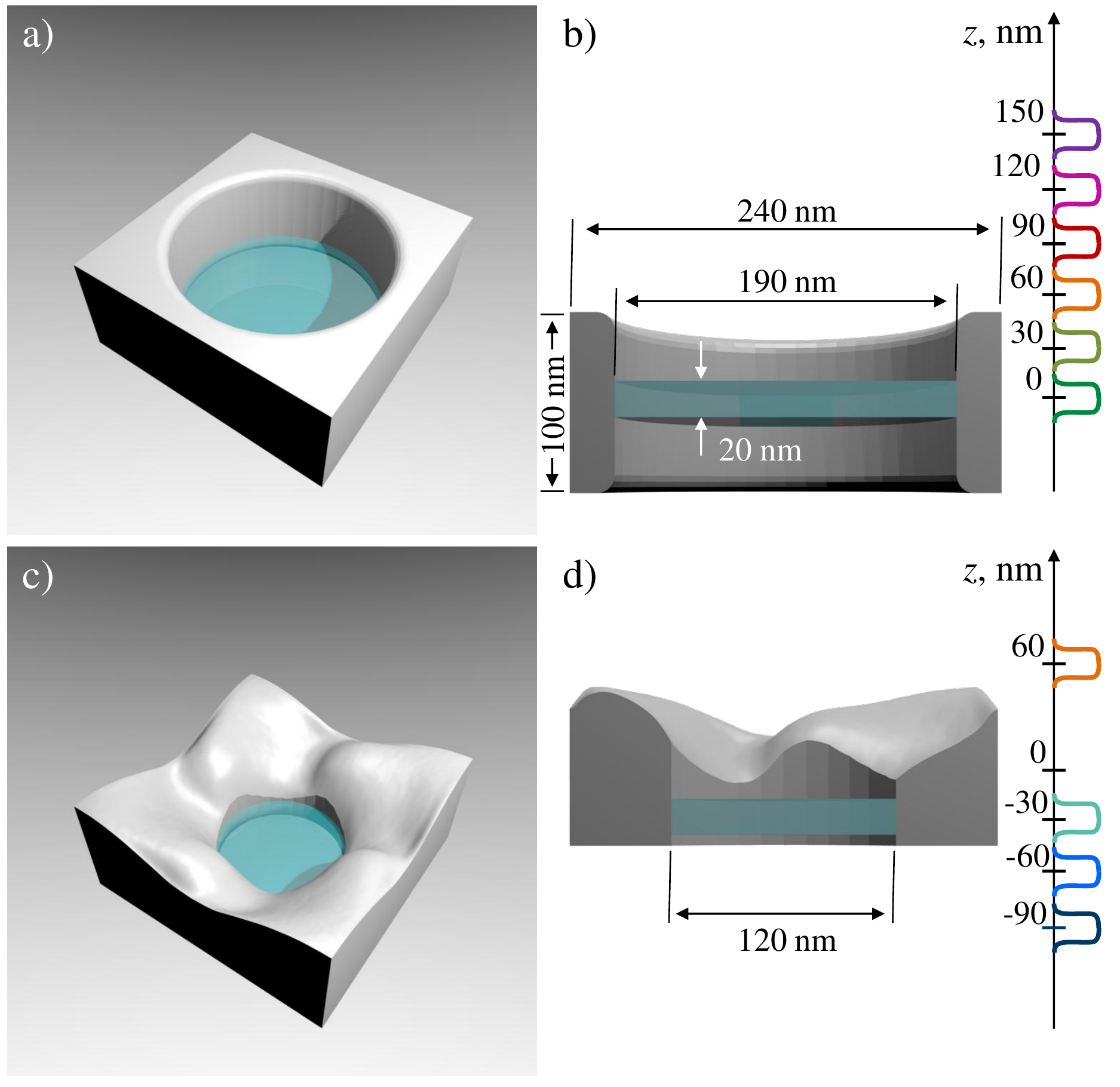}
\caption{Arrays of metal holes with a thin optically active layer: (a) unit cell of the cylindrical hole array; (b) cross section of the cylindrical hole with the layer positions shown on the right; (c) unit cell of the chiral hole array; (d) cross section of the cylindrical hole with the layer positions shown on the right.}
\label{fig:scheme}
\end{figure}

We assume that the metal parts of the hole arrays possess a frequency dependent permittivity $\varepsilon_m(\omega)$ which we take from the experimental reference data for silver \cite{Palik}. The arrays are supposed to be surrounded by a lossless dielectric with a constant refractive index $n_d=1.5$ (permittivity $\varepsilon_d=2.25$). 

For a fair comparison of the OA enhancement by arrays of different hole shape, we performed a careful selection of the arrays that exhibit similar optical performance in the same wavelength range and have identical geometrical and material parameters. These conditions are fulfilled for the arrays of holes in silver film of thickness $w=100$~nm which are schematically presented in Fig.~\ref{fig:scheme}. Both arrays have square lattices with the periods $p=240$~nm, which excludes the visible light diffraction. Their 4-fold rotational symmetry ensures the absence of linear dichroism, i.e., all polarization features in the transmission occur due to the chirality. 
  
The first simple array (Fig.~\ref{fig:scheme}a) consists of cylindrical holes 190 nm in diameter with the metal edges rounded with the radius of 10 nm. For the second array (Fig.~\ref{fig:scheme}c), a complex hole shape resolved by our recent precise microscopic reconstruction of the real fabricated arrays \cite{Kondratov2016} was taken. The shape was rescaled to the period and thickness mentioned above. As we show below, both arrays indeed have very similar light transmission characteristics and, in particular, the same minimum of transmission in the same wavelength range.

The natural optical activity present within the thin layer is accounted via the gyrotropy constant $\alpha$ entering the constitutive relations \cite{1001}:
\begin{equation} 
\begin{array}{l}{\bf D}=\varepsilon_0\varepsilon_d\left({\bf E}+\alpha{\bf
	\nabla}\times{\bf E}\right),\\
{\bf B}=\mu_0\left({\bf H}+\alpha{\bf \nabla}\times{\bf
	H}\right).\end{array}\label{10ad} 
\end{equation}

In a volume of such isotropic chiral material, LCP and RCP waves propagate with the refractive indexes $n_L=n_d+\alpha\varepsilon_d\omega/c$ and $n_R=n_d-\alpha\varepsilon_d\omega/c$ respectively. Accordingly, for a real gyrotropy constant $\alpha$, a layer of thickness $l$ of such material rotates the linear light polarization by an angle  $OA_l=\varepsilon_d\alpha l\omega^2/c^2$. As a quantitative estimate, one can consider the value $\alpha_Q\simeq1.5\cdot10^{-3}$nm. It represents, e.g., the natural optical activity of quartz, a 1 mm layer of which rotates by $\sim30^{\text{o}}$ the polarization of light of the $\lambda=500$~nm wavelength.  

To reveal the spatial selectivity of the chirality sensing, we consider the effects produced by a 20 nm thin optically active layer located differently with respect to the array. The corresponding $\alpha$ profiles and the layer positions used below are illustrated on the right in Figs.~\ref{fig:scheme}b and \ref{fig:scheme}d.

\subsection{Modeling technique}

We use the coordinate system with the $xy$-plane
coinciding with the middle plane of the array of holes perforated in the metal film, and the $z$-axis directed upwards.
The electromagnetic waves are described in terms of magnetic fields.
We consider monochromatic fields and omit the factor $e^{-i\omega t}$ in the following.  

The unit harmonic plane wave
\begin{equation}\label{1ad}
{\bf h}_{in}({\bf r})={\bf H}_{in} e^{-ikz}
\end{equation}
incident on the film from above is supposed to be linearly polarized along the $x$-axis, $H_{in,x}=1$, and $k=n_d\omega /c$.

Below the array, the outgoing transmitted wave ${\bf h}_{tr}({\bf r})$ has the amplitude ${\bf H}_{tr}$ and propagates also against the $z$-axis:
\begin{equation}
	{\bf h}_{tr}({\bf r})={\bf H}_{tr} e^{-ikz}.\label{1ada}
\end{equation}
Knowing the components  $H_{tr,x}$ and $H_{tr,y}$
allows us to express the transmission
amplitudes $t_{R}$ and $t_{L}$ of the RCP and LCP light respectively:
% formula 4ad
\begin{equation} 
	t_{R,L}=H_{tr,x}\mp iH_{tr,y}, \label{4ad}
\end{equation} 
where the upper sign corresponds to $t_{R}$.
The optical chirality parameters,  CD and OA, are then evaluated as: 
\begin{equation}
\label{eq_CD_definition}
CD = \arctan \left(\frac{|t_R| - |t_L|}{|t_R| + |t_L|}\right),
\end{equation}
\begin{equation}
\label{eq_OA_definition}
OA = \frac{1}{2}(\arg t_L - \arg t_R).
\end{equation}
Note that the sign of OA is defined as in the transmission experiments, where positive OA corresponds to the clockwise polarization rotation as seen against the transmission direction. 

The transmittance of the linearly polarized light is expressed as $T=|{\bf H}_{tr}|^{2}$.
Introducing the vector amplitude ${\bf H}_{r}$ of the reflected wave
\begin{equation}\label{1adb}
{\bf h}_{r}({\bf r})={\bf H}_{r} e^{ikz}
\end{equation}
we write the array reflectance as $R=|{\bf H}_{r}|^{2}$.  The light absorption rate is defined as $A=1-R-T$.

We use the finite element method in conjunction with the
eigenmode expansion method (see, e.g., Ref. \cite{1000}) to compute the magnetic fields of the waves that arise as a result of the
diffraction of wave (\ref{1ad}). 
Namely, outside the layer $|z|\le d$ fully embedding the array, the magnetic fields are sought in the form:
\begin{equation}\label{5ad}
{\bf h}^{(+)}({\bf r})={\bf h}_{in}({\bf r})+{\bf h}_{r}({\bf
r})+{\bf h}^{(+)}_{d}({\bf r}),
\end{equation}
for $z\ge d$, and 
\begin{equation}\label{8ad}
{\bf h}^{(-)}({\bf r})={\bf h}_{tr}({\bf r})+{\bf
	h}^{(-)}_{d}({\bf r}),
\end{equation}
for $z\le-d$.
Here the last terms include the diffracted waves:
\begin{equation}\label{6ad}
{\bf h}_{d}^{(\pm)}({\bf r})=\sum_{n,m} {\bf
	H}_{nm}^{(\pm)}e^{i[\pm k_{nm}z+{2\pi}(nx+my)/p]},
\end{equation}
with  
\begin{equation}\label{7ad}
k_{nm}=\sqrt{\varepsilon_d\frac{\omega^2}{c^2}-\frac{4\pi^{2}}{p^{2}}(n^{2}+m^{2})},
\end{equation} 
where ${\text{Im}}(k_{nm})\ge0$.
The summation is performed over the integers $m$ and
$n$, excluding $m=n=0$.

Inside the domain $|z|\le d$, the wave equation for the magnetic field reads as
\begin{equation}\label{11ad}
\frac{1}{\varepsilon}{\bf \nabla}\times{\bf
	\nabla}\times{\bf
	H}=\frac{\omega^2}{c^2}\left({\bf H}+2\alpha{\bf\nabla}\times{\bf H}\right), 
\end{equation}
where $\varepsilon$ takes the values $\varepsilon_m$ and $\varepsilon_d$ inside the metal and dielectric parts respectively, and the gyrotropy constant $\alpha$ is non-zero only inside the optically active dielectric layer. \eqref{11ad} 
is solved by the finite element method within the square structure unit cell with the periodic boundary conditions.

Next, the amplitudes ${\bf H}_{r}$, ${\bf H}_{tr}$ and ${\bf H}_{nm}^{(\pm)}$ are obtained from the matching of the magnetic field computed from \eqref{11ad} and the fields given by \eqref{5ad} at $z=d$ and by \eqref{8ad} at $z=-d$. In the expansion (\ref{6ad}), only the modes are kept, for which the factor $e^{-\text{Im}(k_{nm})d}$ is greater than $10^{-3}$. We have set $d=p/2$ and checked that neither increasing $d$ nor decreasing this small threshold value hardly affects our results.

Controlling the numeric precision becomes critically important as one attempts to reveal the fine polarization features caused by nanosize volumes with weak natural OA.
Solving the problem for the truly achiral cylinder hole array with $\alpha\equiv0$
has indicated the presence of a few millidegrees of parasitic OA and CD due to computational errors.
In order to eliminate this "background", we have exploited the fact that  the effects caused by the layer are proportional to its gyrotropy constant $\alpha$. We have assured that this proportionality holds precisely true up to the values of $\alpha$ exceeding $\alpha_Q$ by more than three orders of magnitude. The modeling results described below were obtained for $\alpha=100\alpha_Q$, for which the erroneous contributions to CD and OA are negligible, and then appropriately renormalized. 

For the final accuracy check, we have verified the features following from the Lorentz reciprocity condition. For 4-fold rotationally symmetric arrays, the latter implies the absence of polarization conversion during reflection and the independence of OA and CD on the side of incidence \cite{PRB2016}. The fact that $xy$-plane is the mirror symmetry plane of the cylindrical hole array determines also that equal OA and CD are induced in this array by the layers located symmetrically with respect to its middle plane. Our  calculations fully reproduced all these facts with the accuracy of the line widths of the curves in the following. 
  
\section{Results}\label{sec:3}

\subsection{Cylindrical hole array}\label{sec:3a}

\begin{figure}%[htbp]
	\centering
	\includegraphics[width=0.9\linewidth]{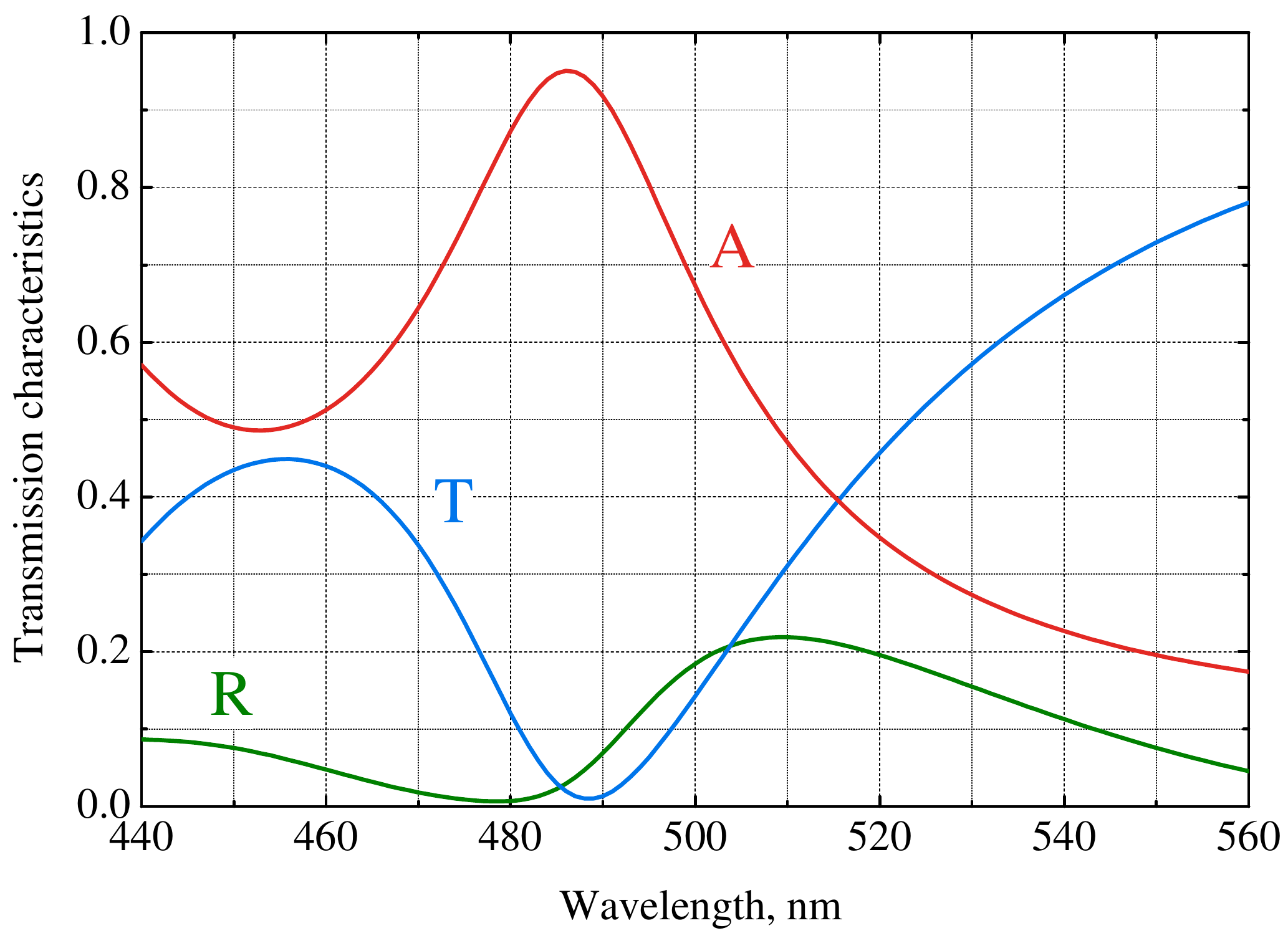}
	\caption{Transmission (T), reflection (R) and absorption (A) spectra of the cylindrical hole array.}
	\label{fig:TRAcyl}
\end{figure}

Modeled transmission of normally incident light through cylindrical hole array has revealed specific spectral dependencies typical for plasmonic structures. As seen in Fig.~\ref{fig:TRAcyl}, both the array transmittance $T$ and reflectance $R$ exhibit minimums at the wavelengths of 488 nm and 479 nm respectively with the minimal transmittance at the 1\% level. The light absorption rate $A$ reaches its well pronounced maximum of 95\% at the 486 nm wavelength. Such spectral behavior is a clear evidence of plasmon resonance.

\begin{figure}%[htbp]
	\centering
	\includegraphics[width=0.9\linewidth]{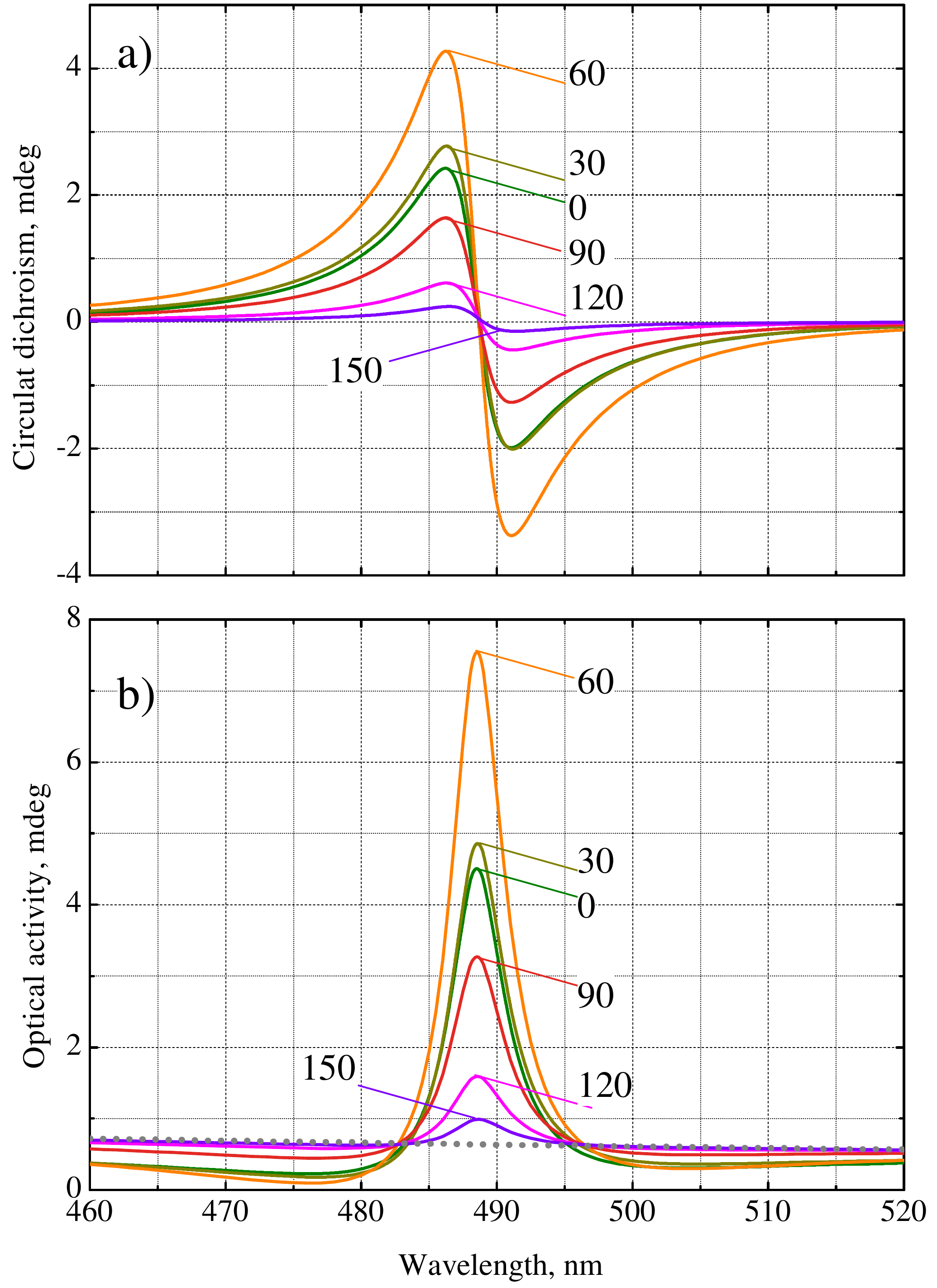}
	\caption{CD (a) and OA (b) of the cylindrical hole array induced by a 20 nm thin optically active layer. The layer positions (in nm) with respect to the array are indicated. Dotted gray line in (b) represents the layer intrinsic OA. The values are normalized for the layer with the natural OA of quartz, $\alpha=\alpha_Q$.}
	\label{fig:CDOAcyl}
\end{figure}

In the presence of a 20 nm thin optically active layer, the array demonstrates specific spectra of resonant optical chirality: narrow peaks of OA and anti-resonant kinks of CD, as shown in Fig.~\ref{fig:CDOAcyl}. The spectral position of the chiral resonance at the 488 nm wavelength is very close to that of the plasmon resonance, while its half-width of only 2.5 nm is almost an order of magnitude smaller than that of the plasmon resonance (about 20 nm). For the wavelengths outside the resonance vicinity, OA quickly drops down to its background value produced by the homogeneous chiral layer on its own, while CD relaxes to zero. 

The particular degree of the resonant OA enhancement demonstrates a very sharp dependence on the chiral layer position. As seen in Fig.~\ref{fig:CDOAcyl}b, the maximum OA exceeds the background level by more than an order of magnitude when the middle of the layer is at 60 nm above the middle plane of the array. Already a 30 nm decrease of this distance lowers OA by a factor of 0.6, while a 30 nm increase halves the peak of OA. One can see from Fig.~\ref{fig:CDOAcyl}a that CD is similarly sensitive to the chiral layer position. Since the effects produced by the chiral layers shifted equally upwards or downwards with respect to the middle plane are identical, only the data for positive layer positions are shown in Fig.~\ref{fig:CDOAcyl}.

\subsection{Chiral hole array}\label{sec:3b}

\begin{figure}%[htbp]
	\centering
	\includegraphics[width=0.87\linewidth]{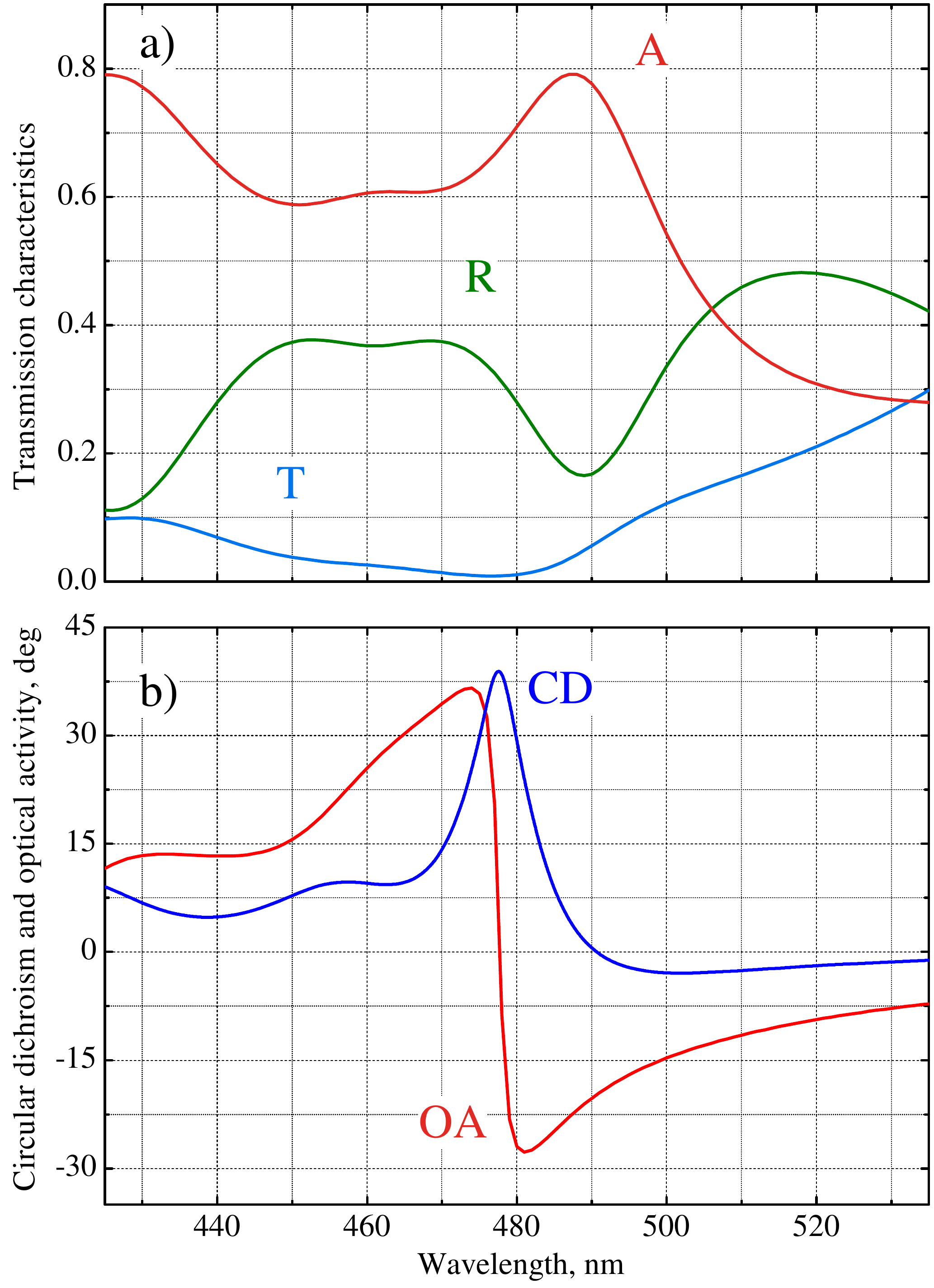}
	\caption{Optical properties of the chiral hole array: (a) transmission (T), reflection (R) and absorption (A) spectra; (b) circular dichroism (CD) and optical activity (OA).}
	\label{fig:dyrkaTRACDOA}
\end{figure}

Transmission and reflection spectra calculated for the array of chiral holes also demonstrate the presence of plasmon resonance. As seen in Fig.~\ref{fig:dyrkaTRACDOA}a, here $T$ drops to the level of about 1\% at the wavelength of 477 nm, while the minimum of $R$ is at 489 nm. The absorption rate $A$ peaks to 79\% at the 487 nm wavelength. The chiral shape of the holes gives rise to remarkably strong optical chirality of the transmitted light: CD drops down to $-40^{\circ}$ thus approaching the extreme minimal possible value of $-45^{\circ}$, while OA varies from about $-35^{\circ}$ to almost $30^{\circ}$ in the narrow spectral range. Such a strong chiral resonance is typical for the holes of this chiral shape and has been observed in the recent experiments \cite{Gorkunov2014} and explained theoretically \cite{PRB2016}.

\begin{figure}%[htbp]
	\centering
	\includegraphics[width=0.9\linewidth]{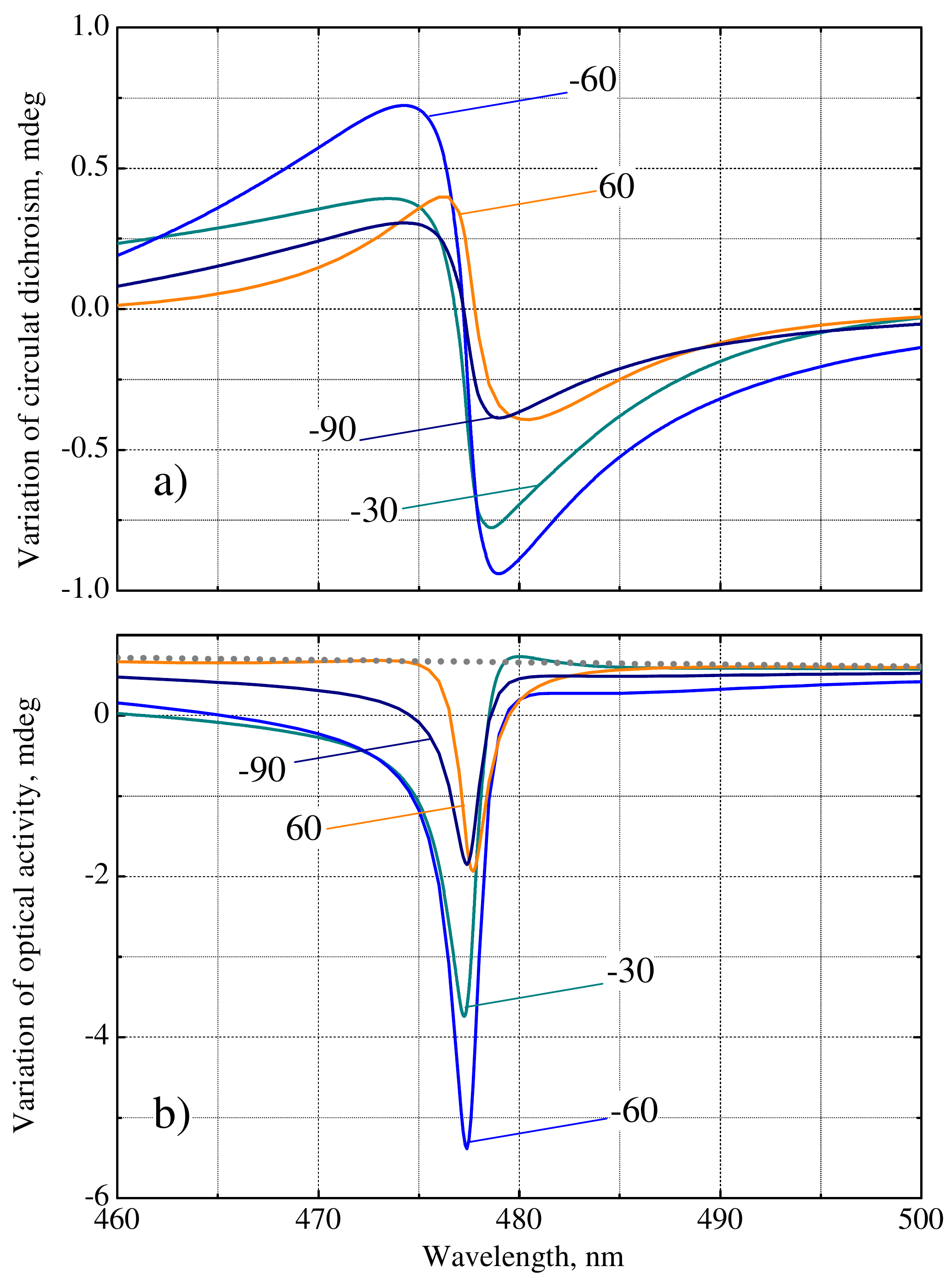}
	\caption{Variation of CD (a) and OA (b) of the chiral hole array induced by a 20 nm thin optically active layer. The layer positions (in nm) with respect to the array are indicated. Dotted gray line in (b) represents the layer intrinsic OA. The values are normalized for the layer with the natural OA of quartz, $\alpha=\alpha_Q$.}
	\label{fig:DCDDOA}
\end{figure}

Adding a 20 nm thin chiral layer causes a small variation of the transmitted light polarization. The corresponding variations of CD and OA are shown in Fig.~\ref{fig:DCDDOA} for different positions of the chiral layer with respect to the hole array. One can see that these spectra are qualitatively similar to those obtained for the cylindrical hole: at the wavelength of the transmission minimum there is a resonant enhancement of the layer-induced OA and an anti-resonant kink of the layer-induced CD. The effect of the optically active layer quickly relaxes to the background levels outside of the resonant range. 

The degree of the chiral sensitivity enhancement here also strongly depends on the position of the chiral layer. In contrast to the cylindrical hole, here the sensitivity is rather asymmetric: it is maximal for the layer attached to the bottom of the structure (relative position $-60$ nm) and noticeably lower when the layer is attached to the top of the structure (relative position $60$ nm). A shift from the $-60$ nm position by some 30 nm substantially affects the observable optical chirality: e.g., OA drops down by three times when the layer is lowered from the $-60$~nm to the $-90$~nm position. 

\section{Discussion}\label{sec:5}

We have performed the full-scale electromagnetic simulations of the complex nanoscale  arrangements comprising plasmonic nanostructures and dielectric materials with natural (molecular) OA. The latter was supposed to be confined to nanosize parts of the dielectric filling, which allowed us to explicitly test the local chiral sensitivity of the structures. 

By comparing Figs.~\ref{fig:CDOAcyl} and \ref{fig:DCDDOA}, one can see that OA and CD acquired by the cylindrical hole array and the variations of OA and CD of the chiral hole array have very much in common. In both cases, the effect of the chiral layer is confined to a narrow spectral range adjacent to the dip of the array transmission. Quantitatively, the range of the changes of CD and OA induced by the chiral layer is almost identical, although in the cylindrical holes this effect occurs on a zero background of the initially achiral transmission, while in the  chiral holes the background consists of an almost extreme optical chirality. 
The chiral sensitivity of both arrays is also similarly selective in space: it is maximal when the layer is attached to one side of the structure, and already a 30 nm displacement of the layer causes a noticeable drop.

\begin{figure}%[htbp]
	\centering
	\includegraphics[width=\linewidth]{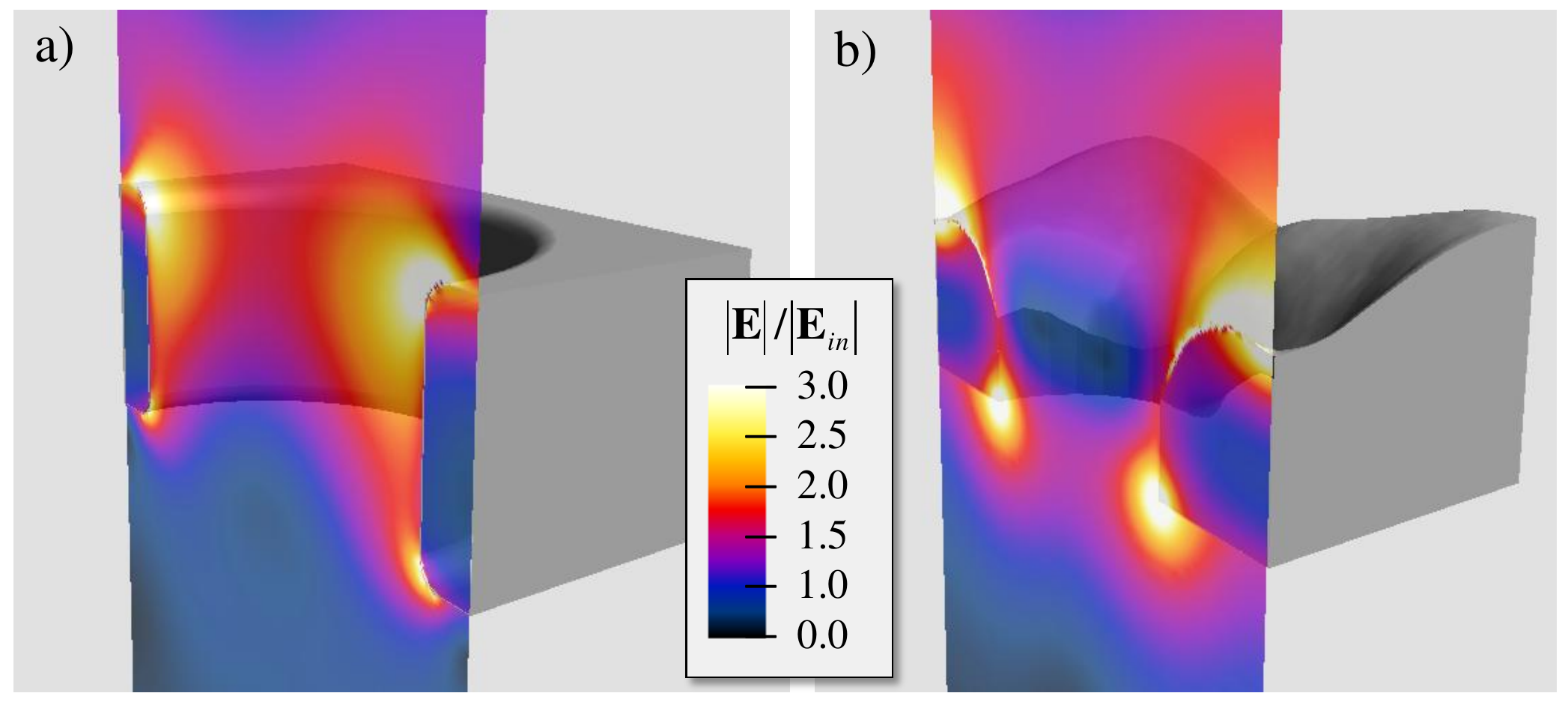}
	\caption{Root mean square of the absolute values of electric field on the $y=0$ slice of the array unit cell normalized by the incident wave amplitude. The light incident from above is linearly polarized along the $x$-axes and its wavelength corresponds to the plasmon resonances of the hole arrays: the cylindrical hole array at the wavelength of 486 nm (a), and the chiral hole array at the wavelength 487 nm (b).}
	\label{fig:fields}
\end{figure}

This remarkable similarity of the spectral and spatial sensitivity of the arrays of different types is related to the fact that they both host plasmon resonances which dominate their optical properties. The pronounced peaks of the light absorption seen in Figs.~\ref{fig:TRAcyl} and \ref{fig:dyrkaTRACDOA}a allowed us to determine the plasmon resonant wavelength values as 486 nm and 487 nm for the cylindrical hole and chiral hole arrays respectively. We performed the finite-difference time-domain modeling of both arrays with SPEAG SEMCAD X FDTD solver and calculated the corresponding near field distributions shown in Fig.~\ref{fig:fields}. 

One can  see in Fig.~\ref{fig:fields}a that the plasmon resonance of the cylindrical hole has two hot-spots: at the top and at the bottom of the hole. Although the hot-spots are excited differently when the light is incident from one side (e.g. from the above as in Fig.~\ref{fig:fields}), due to the mentioned symmetry reasons, the effect of the chiral layer is equally maximal when it is placed near either of them. Apparently, the reciprocity of Maxwell's equations determines the exact equality of the observables in these two cases. At the top, the layer is subjected to higher fields, but its effect on the transmission is attenuated while being transfered to the exit. On the other side near the bottom hot-spot, the fields are weaker, but the layer immediately contributes to the transmitted plane wave. 
In the chiral hole, as seen in Fig.~\ref{fig:fields}b, the  hot-spots are also localized near the top and bottom film surfaces. Here, however, their excitation is much more symmetrical. Accordingly, the bottom hot-spot has much stronger effect on the transmission and a chiral layer near it produces the largest variation of the transmitted light polarization. 

Therefore, we conclude that the main mechanism behind the sensitivity of the arrays to the chiral environment is the plasmonic field enhancement. This effect has been analyzed only in the simplest geometries so far ~\cite{Govorov2010,Zhang2013}. Note that CD acquired by a metal sphere from an optically active molecule was evaluated in Ref.~\cite{Govorov2010} as proportional to $(\omega_{\text{plasmon}}-\omega)^{-1}$, i.e., with the anti-resonant frequency dispersion, which is in-line with the CD spectra obtained above. 

An important conclusion following from our simulations is the absence of a notable advantage of chiral complex-shaped structures. Formally, this does not contradict the recent works on the chiral sensing with the structures having strongly chiral near field distributions (see e.g. Ref.~\cite{Schaeferling2014}). The concept of the field chirality that initiated the research on plasmonic superchiral fields, was naturally aimed at enhancing molecular CD. We note, however, that the absence of molecular CD in the visible is a common feature of the majority of organic molecules and biological objects. Instead, one has to optimize the techniques for enhancing the effects of molecular OA. As we have seen, an array of the simplest cylindrical holes has no real disadvantages here. At the same time, one can possibly exploit the positive impact of the near field chirality for amplifying the molecular CD in the ultraviolet range using, for instance, aluminum nanostructures that support plasmon resonances at such frequencies. 

Finally, we notice that due to the broad availability of accurate optical CD detecting techniques, practically all works on plasmon enhanced chirality are focused on CD calculation and measurement. As we have shown, plasmon enhanced OA can noticeably exceed CD and we suggest that developing OA-sensitive techniques can be of advantage for the future plasmon-based chirality sensorics.

\vspace{1cm}

\textbf{Funding.} Russian Science Foundation (project 14-12-00416).

\textbf{Acknowledgments.} We are grateful to A.~A.~Ezhov for valuable discussions and criticism.

\end{document}